\documentclass[12pt]{article}
\usepackage{graphicx, amsmath, amssymb}
\usepackage[numbers, sort&compress]{natbib}
\usepackage{fullpage}

\newcommand{\W}{\mathbf{W}} 
\newcommand{\Q}{\mathbf{Q}}
\newcommand{\beq}{\begin{equation}}
\newcommand{\eeq}{\end{equation}}
\newcommand{\arr}{\rightarrow}
\newcommand{\me}[3]{\langle #1 | #2 | #3 \rangle}

\newcommand{\ket}[1]{| #1 \rangle}
\newcommand{\average}[1]{\langle #1 \rangle}

\newcommand{\s}{\sigma}
\newcommand{\Z}{\mathcal{Z}}
\newcommand{\AP}{\text{AP}}

\title{Statistical Physics of Evolutionary Trajectories \\ on Fitness Landscapes}
\author{Michael Manhart$^1$ and Alexandre V. Morozov$^{1,2}$\footnote{Corresponding author: \texttt{morozov@physics.rutgers.edu}} \\ 
	\small{\emph{$^1$ Department of Physics and Astronomy, Rutgers University, Piscataway, NJ 08854}} \\
	\small{\emph{$^2$ BioMaPS Institute for Quantitative Biology, Rutgers University, Piscataway, NJ 08854}}
	}
\date{}

\begin{document}

\maketitle

\begin{abstract}
Random walks on multidimensional nonlinear landscapes are of interest in many areas of science and engineering. In particular, properties of adaptive trajectories on fitness landscapes determine population fates and thus play a central role in evolutionary theory.  The topography of fitness landscapes and its effect on evolutionary dynamics have been extensively studied in the literature. We will survey the current research knowledge in this field, focusing on a recently developed systematic approach to characterizing path lengths, mean first-passage times, and other statistics of the path ensemble. This approach, based on general techniques from statistical physics, is applicable to landscapes of arbitrary complexity and structure. It is especially well-suited to quantifying the diversity of stochastic trajectories and repeatability of evolutionary events. We demonstrate this methodology using a biophysical model of protein evolution that describes how proteins maintain stability while evolving new functions.
%The evolutionary trajectories and fates of populations are matters of central interest in evolutionary theory.  A common approach to modeling these systems uses the idea of a fitness landscape over which populations adapt.  Thus much research has been devoted to studying the topography of these landscapes, especially how they depend on the underlying molecular processes of proteins and DNA, and how evolutionary trajectories proceed on these landscapes.  We will survey the current research knowledge on these topics, and then discuss some general techniques from statistical physics for studying trajectories on these landscapes.  We will focus especially on characterizing the diversity of evolutionary trajectories.  Finally, we will explore these tools on a biophysical model of protein evolution.
\end{abstract}

%%%%%%%%%%%%%%%%%%%%%%%%%%%%%%%%%%%%%%%%%%%%%%%%%%%%%%%%%%%%%%%%%%%%%%%%%%%%%%%%%%%%%%%%%%%%%%%%%%%%

\section{Introduction}

     Random walks on networks are ubiquitous in nature.  For example, consider proteins, the macromolecules that carry out a myriad of chemical and mechanical functions inside a cell~\cite{Nelson2007}.  Each protein is a chain of amino acids chemically bonded to make a linear polypeptide~\cite{Creighton1992},
%.  There are 20 naturally occurring amino acids, 
and the sequence of amino acids determines the protein fold --- a compact 3D conformation which has the minimum free energy.
%by making favorable energetic contacts between amino acids that are close in space, but not necessarily close along the amino acid sequence.
% An aspect of protein energetics is hydrophobicity: since proteins fold in water (or, more precisely, in complex solutions found within cells), conformations with low free energies tend to have hydrophobic amino acids on the inside and hydrophilic amino acids on the outside, in contact with water molecules.
Unlike random heteropolymers, naturally-occuring proteins have unique folds that they achieve robustly and, in many cases, rapidly (on the time scales of micro- or milliseconds) starting from arbitrary unfolded conformations~\cite{Finkelstein2002}.

     Proteins are produced in the unfolded state inside the cell and have to fold before they can function.  Protein conformations are often represented by sets of dihedral angles (torsion angles defined by three bond vectors connecting four atoms~\cite{Finkelstein2002}).
%; in this approximation, all bond lengths and bond angles are considered to be fixed.
Although in general the values of dihedral angles are continuous (i.e., atoms can freely rotate around dihedral bonds), they are typically discretized in protein structure prediction algorithms.  In this case, protein folding can be viewed as a random walk on a network with connectivity defined by the move set --- a set of instructions for changing the dihedral angles in each step.

     The network is very high-dimensional.  For example, for a relatively small protein with $L=100$ amino acids, $2$ dihedral angles per amino acid, and $10^\circ$ dihedral angle increments, there are $36^{200}$ possible conformations.  With a simple move set that updates one angle at a time, each node is connected to $35$ neighbors.  In such a large space, how can a protein reach its unique folded shape on reasonable time scales?  This problem is known as the Levinthal paradox~\cite{Levinthal1968}, and key to its resolution is the idea of the protein folding landscape~\cite{Bryngelson1995, Dill2012}.  In our example above, each protein conformation has a free energy which is a function of $200$ dihedral angles, forming a landscape over the network.  The free energy values at a node and its neighbors determine rates of transition between nodes, e.g., according to the Metropolis algorithm~\cite{Metropolis:1953}.  This landscape is believed to have a global funnel shape, which allows the protein to efficiently find its folded structure through incremental moves without searching the entire space~\cite{Dill2012}.
     
     This picture generalizes to many other search problems on landscapes and networks in which each node, corresponding to a discrete (or discretized) state of the system, can be assigned a value of the objective function which sets the transition rates.  As with protein folding, a major question is how the landscape topography and the move set determine the dynamics.  For example, an important quantity of interest is the mean first-passage time (to the global minimum on the protein folding landscape, for example), which should be minimal in optimized algorithms~\cite{Tolkunov2012}.

% which employ stochastic sampling techniques such as Metropolis Monte Carlo (MC) search~\cite{Metropolis:1953}, simulated annealing~\cite{Kirkpatrick:1983}, or genetic algorithms~\cite{Goldberg:1989} to find the minimum free energy conformation starting from random points on the landscape. 
%Each node on the network is assigned a value according to the continuous free energy function into which the network is embedded.     
%The terms in the function account for electrostatic interactions, hydrogen bonds, van der Waals contacts, short-range interatomic repulsion, etc.

%     In some cases, there is no embedding multi-dimensional potential surface and the problem is fundamentally discrete.
     The effect of landscape topography on dynamics is of particular importance in evolutionary theory, the study of how populations of organisms change over time through mutation and natural selection~\cite{Darwin1859}.  The genotype (genetic state) of an organism is represented by a sequence $\s$ of letters drawn from an alphabet of size $k$.  The sequence may represent nucleotides in genomic DNA ($\{\mathsf{A}, \mathsf{C}, \mathsf{G}, \mathsf{T}\}$, $k=4$), amino acids in a protein ($k=20$), or the presence/absence of a mutation at several genes across the genome ($k=2$).  Assuming a fixed number $L$ of sites in each sequence, the space of all $k^L$ possible sequences represents a network with sequence nodes connected to each other if they differ by a mutation at a single site~\cite{MaynardSmith1970}.  For simplicity we neglect recombination between sequences and insertion/deletion of sites which would redefine network connectivity and in some cases the total number of nodes.  Each sequence $\s$ can be assigned a fitness value $\mathcal{F}(\s)$, defined as the survival probability of an individual with that sequence~\cite{Gillespie2004}. The resulting construct is called a fitness landscape or, more precisely, a genotypic fitness landscape~\cite{Wright1932}.  Just as the folding landscape's structure is key to a protein's ability to reach its folded state efficiently, the fitness landscape is key to understanding how complex biological structures, such as bacterial flagella or the human eye, can arise through random, incremental mutations~\cite{MaynardSmith1970}.
     
\subsection{Evolutionary dynamics}

     In general, individuals in a population will have different sequences, occupying a distribution of points on the fitness landscape. However, in the limit $u \ll (N \log N)^{-1}$~\cite{Champagnat2006a, Champagnat2006b}, where $u$ is the mutation rate (defined as the probability of mutation per site per generation) and $N$ is the effective population size~\cite{Crow1970, Gillespie2004}, new mutations arise individually and either fix in the population or disappear from it on time scales that are short compared with the times between successive mutations~\cite{Kimura1983, Lassig2007, Manhart2012}.  Thus the population is monomorphic and, apart from short transition periods, occupies a single node in sequence space.  The stochastic process of a new mutant appearing and fixing in the population is known as substitution, and the substitution rate from sequence $\s$ to $\s'$ is given by~\cite{Kimura1983}
     
\beq
\me{\s'}{\W}{\s} = Nu \phi(s),
\eeq

\noindent where $Nu$ is the total number of new mutations per generation ($Nu \ll (\log N)^{-1} < 1$), and $\phi(s)$ is the probability of a single $\s'$ mutant fixing in a population of $\s$ when the selection coefficient is $s = \mathcal{F}(\s')/\mathcal{F}(\s) - 1$ ($s > 0$ for beneficial mutations, $s < 0$ for deleterious ones).

     The exact form of the fixation probability $\phi(s)$ depends on the underlying population dynamics.  However, it is common to consider the strong-selection weak-mutation (SSWM) limit in which beneficial mutations always fix and deleterious mutations always get eliminated~\cite{Gillespie1984}.  This approximation is accurate for $|s| \gg 1$ and $N \gg 1$.  Similar to zero-temperature Monte Carlo, the population can only undergo substitutions that increase fitness, and all substitutions occur with the same rate $Nu$ (since for $|s| \gg 1$, $\phi(s) \approx 1$ when $s > 0$ and $\phi(s) \approx 0$ when $s < 0$).  Thus adaptation follows trajectories on the landscape along which fitness increases monotonically.
     
     Other approximations may be more appropriate in different circumstances~\cite{Wakeley2005}.  For example, when $1 \ll N|s| \ll N$, $\phi(s) \approx s$ for $s > 0$ and $\phi(s) \approx 0$ for $s < 0$.  Thus deleterious mutations always get eliminated as before, but beneficial mutations fix at the rate $Nus$.  More generally, while the true dynamics of real populations may involve interference between multiple simultaneous mutations and other complexities~\cite{Desai2007}, simplified evolutionary dynamics are useful when our main objective is to understand the role of the fitness landscape in constraining evolution.

%%%%%%%%%%%%%%%%%%%%%%%%%%%%%%%%%%%%%%%%%%%%%%%%%%

\subsection{Epistasis}

% epistasis
     The most basic aspect of fitness landscape topography is known as \emph{epistasis}.  Let the sequence $\s$ be $\s^1 \s^2 \ldots \s^L$, where $\s^\mu$ is the letter at site $\mu \in \{1, \ldots, L\}$.  In general the fitness function $\mathcal{F}(\s)$ cannot be decomposed into a sum of independent contributions from each site $\mu$.  This means that the fitness effect of a mutation at a given site may depend on the state of other sites.  If this is true, the sites will be correlated, which can be thought of as a coupling or interaction between the sites, similar to a Hamiltonian for a system of interacting particles.
%For free particles, the Hamiltonian consists of additive contributions from each particle, which means their states are uncorrelated.  Interacting (correlated) particles, however, give rise to terms in the Hamiltonian that cannot be decomposed into independent contributions.

\begin{figure}[t!]
\begin{center}
\includegraphics[scale=0.5]{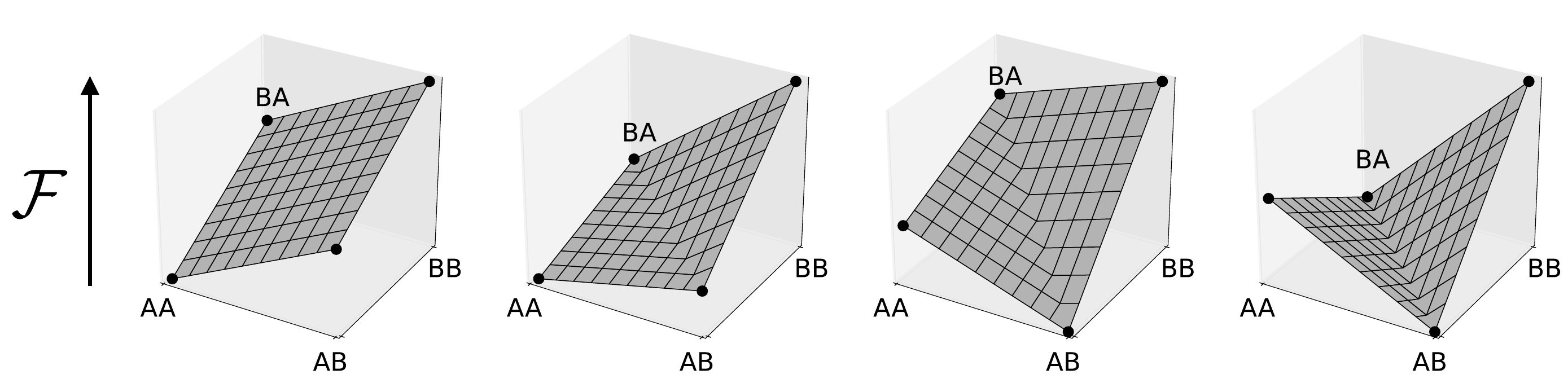}
\end{center}
\caption{The four qualitative types of epistasis for a two-letter, two-site model.  From left to right: no epistasis, where each mutation has the same additive effect on fitness regardless of the other site, yielding a linear landscape; magnitude epistasis, where the magnitude (but not the sign) of the fitness effect of a mutation depends on other sites; sign epistasis, where the sign of a mutation's fitness effect (beneficial or deleterious) depends on other sites; reciprocal sign epistasis, where multiple instances of sign epistasis can lead to multiple local fitness maxima.}
\label{fig:epistasis}
\end{figure}

     Epistasis is precisely this interactive coupling in the context of genotypic sequences.  We categorize types of epistasis according to the qualitative differences in the fitness effects of mutations.  We summarize the four possible cases using a two-letter, two-site model in Fig.~\ref{fig:epistasis} in which sequence $\mathsf{AA}$ evolves into $\mathsf{BB}$, which has the highest fitness.  In the first case on the left of Fig.~\ref{fig:epistasis}, there is no epistasis: the fitness effect of $\mathsf{A} \to \mathsf{B}$ substitution at site 2 is the same regardless of the state of site 1, and vice versa.  Thus the fitness can be decomposed into a sum of additive contributions from each site: $\mathcal{F}(\s) = \mathcal{F}_1(\s^1) + \mathcal{F}_2(\s^2)$, i.e., the landscape is linear in sequence space.  Under SSWM evolutionary dynamics, both trajectories from $\mathsf{AA}$ to the global maximum at $\mathsf{BB}$ are accessible.
     
     In the second case of Fig.~\ref{fig:epistasis}, the fitness effect of $\mathsf{A} \to \mathsf{B}$ at site 2 differs in magnitude but not in sign depending on whether site 1 has $\mathsf{A}$ or $\mathsf{B}$.  This situation is known as \emph{magnitude epistasis}~\cite{Carneiro2010, Weinreich2005}.  Magnitude epistasis does not completely block any trajectories, although it may affect quantitative aspects of dynamics such as adaptation times.  Note that there are two kinds of magnitude epistasis: one in which the fitness benefit of a mutation is enhanced by the presence of other mutations, and the ``diminishing returns'' case in which the fitness benefit is decreased by other mutations.
     
     The third case of Fig.~\ref{fig:epistasis} shows how the $\mathsf{A} \to \mathsf{B}$ substitution at site 2 can have opposite effects on fitness depending on the state of site 1: it is deleterious if $\s^1 = \mathsf{A}$, but beneficial if $\s^1 = \mathsf{B}$.  Since the sign of the fitness effect depends on the other site, the situation is known as \emph{sign epistasis}. Sign epistasis can significantly affect accessibility of genotypes on the landscape by blocking trajectories: under SSWM dynamics, the trajectory $\mathsf{AA} \to \mathsf{AB} \to \mathsf{BB}$ is unavailable since it requires a deleterious substitution.  When sign epistasis exists at multiple sites, it is known as \emph{reciprocal sign epistasis}, as shown in the fourth case of Fig.~\ref{fig:epistasis}. Reciprocal sign epistasis is a necessary condition for the existence of multiple local maxima~\cite{Poelwijk2011}.
%These cases are straightforwardly generalized to higher-dimensional sequence spaces with additional sites and letters.
As the examples in Fig.~\ref{fig:epistasis} show, epistasis underlies landscape ruggedness that can constrain evolutionary trajectories.  Thus the existence and nature of epistasis is of a prime interest in evolution.

\subsection{Measures of landscape ruggedness and accessibility}

     Numerous measures have been proposed to quantify ruggedness of fitness landscapes and accessibility of evolutionary trajectories (summarized in Ref.~\citep{Szendro2013}).  One commonly-used measure is the number of local fitness maxima, which is indicative of the presence and type of epistasis~\cite{Poelwijk2011}: more local maxima indicates a more rugged or epistatic landscape.  For binary alphabets, deviations of the fitness function from linearity can be quantified by fitting a linear function and calculating the sum of squares of residuals, known here as a roughness parameter~\cite{Szendro2013}.  A more local measure of ruggedness can be obtained by considering all pairs of sites and all pairs of possible letters at those sites, and then determining the sub-landscape for each combination like those shown in Fig.~\ref{fig:epistasis}.  Each sub-landscape can then be classified into one of the four epistasis types.
     
     Other measures consider accessibility and other properties of the trajectories themselves, especially those leading to the global fitness maximum.  For example, without epistasis all trajectories to the global maximum from any other point on the landscape are accessible under SSWM dynamics, but with sign epistasis some trajectories become blocked.  The distributions of trajectory times and lengths are also important for understanding the effect of landscape ruggedness on adaptation~\cite{Flyvbjerg1992, Traulsen2007, Kryazhimskiy2009, Franke2011, Manhart2013}.
     
%%%%%%%%%%%%%%%%%%%%%%%%%%%%%%%%%%%%%%%%%%%%%%%%%%
     
\subsection{Diversity of evolutionary trajectories and evolutionary determinism}

     Landscape ruggedness is especially relevant in its effect on the predictability of evolution, a question of paramount importance in evolutionary theory~\cite{Lobkovsky2012}.  If ``life's tape'' could be replayed, would we see a completely different outcome because evolution is a largely stochastic phenomenon, or are accessible evolutionary trajectories so constrained that the outcome would have been the same or recognizably similar~\cite{Gould1990}?  Discussing this question in general entails many issues, including environmental conditions, initial conditions, and details of population dynamics.  More specifically, one can focus on the diversity of transition pathways in evolving from an ancestral state to a particular descendant state or set of states.
     
     One assessment of this diversity is simply counting the number of distinct accessible trajectories connecting an initial sequence $\s$ with a final sequence $\s'$.  For example, Weinreich and co-workers found that only a small fraction of all trajectories from wild-type \emph{E. coli} to a strain resistant to antibiotics was accessible to adaptive walks~\cite{Weinreich2006, Poelwijk2007}.  In another approach, Koonin and co-workers devised a measure called mean path divergence to quantify the diversity more precisely~\cite{Lobkovsky2011, Lobkovsky2012}:

\begin{equation}
\mathcal{D} = \sum_{\varphi_1 \ne \varphi_2} d(\varphi_1, \varphi_2) p(\varphi_1) p(\varphi_2),
\label{eq:mpd_Koonin}
\end{equation}

\noindent where the sum is over all pairs of distinct paths in an ensemble, $p(\varphi)$ is the probability of path $\varphi$, and $d(\varphi_1, \varphi_2)$ is the distance between paths $\varphi_1$ and $\varphi_2$, defined as the average of the shortest Hamming distances between each point $\s_1$ on path $\varphi_1$ and all points on path $\varphi_2$, and vice versa:

\begin{equation}
d(\varphi_1, \varphi_2) = \frac{1}{\mathcal{L}[\varphi_1] + \mathcal{L}[\varphi_2] } \left( \sum_{\s_1 \in \varphi_1} h(\s_1, \varphi_2) + \sum_{\s_2 \in \varphi_2} h(\s_2, \varphi_1) \right),
\end{equation}

\noindent where $\mathcal{L}[\varphi]$ is the length (number of steps) of path $\varphi$, and $h(\s_1, \varphi_2)$ is the shortest Hamming distance between $\s_1$ and all points $\s_2 \in \varphi_2$.
The divergence therefore captures not only how many paths are available, but weighs them by their spatial proximity.
Other measures of path diversity, such as path entropy and the distribution of path lengths and times~\cite{Manhart2013}, are discussed later in this article.

%%%%%%%%%%%%%%%%%%%%%%%%%%%%%%%%%%%%%%%%%%%%%%%%%%

\subsection{Model and empirical landscapes}

% simple models of landscapes (examples)
     %Since experimental information on fitness landsapes is restricted to relatively few genotypes, most studies of fitness landscapes are based on simplified theoretical models.
     A few simple models have traditionally dominated theoretical studies of fitness landscapes and evolutionary trajectories. These models serve as useful null hypotheses or limits of more complex scenarios;
%in addition, direct characterization of empirical fitness landscapes is made challenging by the small number of genotypes considered in experimental studies.
they generally consider sequences with binary alphabets, in which case sequence space is a unit hypercube.  Without attempting to account for every landscape proposed in the literature, we will discuss and motivate several more popular choices. In Kauffman's NK model~\cite{Kauffman1989, Kauffman1993}, each of the $L$ sites in the gene (or genes in the genome) interacts with $K$ other sites chosen by random sampling. The fitness of genotype $\s$ is given by

\begin{equation}
\mathcal{F}(\s) = \sum_{\mu=1}^L f_\mu(\s^\mu,\s^{n_1(\mu)}, \dots, \s^{n_K(\mu)}),
\end{equation}

\noindent where $n_1(\mu), \dots, n_K(\mu)$ are interaction partners of site $\mu$. The single-site fitnesses $f_\mu$ are obtained by sampling from a continuous distribution; each combination of $2^{K+1}$ possible states of the argument corresponds to an independent sampling. When $K=0$, the NK landscape becomes fully additive and thus non-epistatic. Because in this limit the landscape is smooth and has a single maximum, it is sometimes called the ``Mount Fuji'' model~\cite{Aita2000}. The amount of epistasis, or landscape ruggedness, can be tuned by increasing $K$ to the maximum value of $L-1$. With $K = L-1$, the fitnesses of different sequences are uncorrelated; this model is called the ``House of Cards''~\cite{Kingman1978} due to the unpredictable fitness effects of mutations.  Since closely-related genotypes will realistically have correlated fitness, this limit serves mainly as a null model.  Various properties of the NK model are known:~\cite{Kauffman1989, Flyvbjerg1992, Rokyta2006, Kryazhimskiy2009}  for example, in the $K = L-1$ limit the average number of local maxima is $k^L/(L(k-1) + 1)$ for any alphabet size $k$~\cite{Szendro2013}.

     Another class of models starts from a non-epistatic landscape and adds noise to it. For example, in the ``rough Mount Fuji'' model~\cite{Aita2000}, sequence $\s_0$ is arbitrarily picked as the global maximum and the fitness of sequence $\s$ is given by

\begin{equation}
\mathcal{F}(\s) = \eta (\s) - \theta d(\s, \s_0),
\end{equation}

\noindent where $d(\s, \s_0)$ is the Hamming distance between sequences $\s$ and $\s_0$, $\theta$ is the parameter which controls the slope of the smooth part of the landscape, and $\eta (\s)$ is a zero-mean random variable sampled independently for each sequence $\s$. The ruggedness of the landscape is controlled by the ratio of $\theta$ and the standard deviation of the distribution from which the random variables $\eta (\s)$ are sampled.

     The landscapes described above are dominated by selection. Another approach to evolution is based on the neutral theory, which postulates that the majority of mutations have either no phenotypic effect (i.e., are selectively neutral) or are strongly deleterious and thus removed from the population~\cite{Kimura1983}. This picture leads to evolution on a neutral network where all viable nodes have the same fitness~\cite{Nimwegen1999, MaynardSmith1970}. With some probability, a viable individual can acquire a lethal mutation from which it cannot recover, and will disappear from the population. Thus the population as a whole can only make transitions between viable, selectively neutral states. Evolution on such a landscape is reminiscent of the percolation problem~\cite{Stauffer1994}: each node is assigned fitness 1 with probability $p$ and fitness 0 with probability $1-p$, independent of the other nodes~\cite{Franke2011}.

% experiments
     Due to the enormous number of sequences involved, the structure of fitness landscapes is difficult to probe experimentally. Typically, only a small number of sites is studied (4 to 9, summarized in Ref.~\citep{Szendro2013}), and at those sites, only a subset of all possible mutations is introduced, resulting in fitness measurements for ${\cal O}(10^2)$--${\cal O}(10^3)$ different genotypes. In addition, because genotype survivability is not directly accessible in experiments, proxy measures of fitness are employed, such as growth rates and antibiotic resistance.  Although these empirical studies can be used to probe the local structure of the landscapes, they are insufficient for analyzing the global properties of adaptive trajectories because adaptation may involve mutations outside of the experimentally-probed subset.
     
     Nevertheless, many studies have attempted to characterize empirical landscapes in terms of their epistatic features, accessibility, and correspondence to theoretical models~\cite{Mammano2000, Weinreich2006, Poelwijk2007, Carneiro2010, Franke2011, Lobkovsky2011, Szendro2013}.  For example, magnitude and sign epistasis have been observed, as well as significant constraints on evolutionary trajectories.
%Models such as the NK model have been fit to data to infer values of $K$.
One general finding of such studies is that empirical landscapes include some epistasis, but are far from the House of Cards regime in which all fitness values are completely uncorrelated~\cite{Carneiro2010}.  The emerging picture is closer to the rough Mount Fuji model, which includes a limited amount of epistasis around a mostly linear landscape~\cite{Szendro2013}.

%%%%%%%%%%%%%%%%%%%%%%%%%%%%%%%%%%%%%%%%%%%%%%%%%%%%%%%%%%%%%%%%%%%%%%%%%%%%%%%%%%%%%%%%%%%%%%%%%%%%

\section{Statistical physics of stochastic paths}

     Analytical treatments of evolutionary dynamics on fitness landscapes are typically restricted to uncorrelated or highly symmetric models, such as those previously described.  Simulations, meanwhile, can suffer from numerical inaccuracy and be computationally expensive when rare events are considered.  More systematic tools are necessary, especially tools that directly address statistical properties of stochastic paths that are relevant for understanding the diversity of evolutionary pathways.

     Physics and chemistry have long grappled with similar problems in the field of reaction rate theory~\cite{Hanggi1990}, which studies rare transitions between metastable states that model phenomena ranging from protein folding~\cite{Finkelstein2002} to chemical reactions~\cite{Bolhuis2002}.  In these systems, quantities of interest include not only mean first-passage times and reaction rates but also the spatial distribution of transition paths and identification of kinetic bottlenecks.
     
     Transition state theory is a well-known approach to these problems; however, it relies on the existence and \emph{a priori} identification of key transition states~\cite{Hanggi1990}.  A more recent development has been transition path sampling~\cite{Dellago1998, Bolhuis2002, Dellago2003, Hummer2004, Harland2007, Mora2012}, in which paths are directly sampled via Monte Carlo to estimate their statistical properties.  Similar methods have been used in phylogenetic analysis of protein sequences~\cite{Robinson2003, Rodrigue2005, Rodrigue2006, Choi2007, Rodrigue2009}.
These techniques are based on a finite sample of paths and do not provide natural cutoffs for the size of the path ensemble, which may lead to noisy estimates of various path statistics. Another technique, called transition path theory~\cite{E2006, Metzner2006, Metzner2009, Noe2009, E2010}, relies on explicit solutions to the backward equation. This approach, though more systematic, does not directly address the diversity of paths.

     Here we discuss a general statistical mechanics treatment of stochastic paths that provides many useful tools for analyzing evolutionary models and other stochastic processes~\cite{Manhart2013}.  A semi-Markov process (i.e., continuous-time random walk~\cite{Weiss1994}) on the state space $\mathcal{S}$ consists of discrete jumps between states and continuous-time waiting within states; the jump process is memoryless, but the waiting process need not be.  Thus the process is defined by a set of jump probabilities, $\me{\s'}{\Q}{\s}$ for the jump $\s \arr \s'$ ($\s,\s'\in\mathcal{S}$) and waiting time distributions $\psi_\s(t)$, which is the PDF of waiting exactly time $t$ in state $\s$ before jumping out.  Assume $\psi_\s(t)$ has finite mean $w(\s)$ for all $\s \in \mathcal{S}$.  For Markov processes with memoryless waiting, $\psi_\s(t) = e^{-t/w(\s)}/w(\s)$.  Non-exponential $\psi_\s(t)$ can also arise due to coarse-graining of Markov processes~\cite{Klafter1980, Redner2001, Maes2009}.  The space $\mathcal{S}$ equipped with the jump matrix $\Q$ defines a network with directed, weighted edges.
     
     Define a trajectory as a path through state space $\varphi = \{\s_0, \s_1, \ldots, \s_\ell\}$ combined with a set of intermediate waiting times $\{t_0, t_1, \ldots, t_{\ell-1}\}$, where $t_i$ is the waiting time in $\s_i$.  We consider the trajectory finished once it reaches the final state $\s_\ell$, so we do not count the waiting time in that state.  The probability functional of starting in the initial state $\s_0$ and completing the path $\varphi$ no later than time $t$ is
     
\beq
\Pi[\varphi, t] = \pi(\s_0) \left( \prod_{i=0}^{\ell-1} \me{\s_{i+1}}{\Q}{\s_j} \right) \left( \prod_{i=0}^{\ell-1} \int_0^\infty dt_i ~\psi_{\s_i}(t_i) \right) \Theta\left(t - \sum_{i=0}^{\ell-1} t_i\right),
\label{eq:full_path_prob}
\eeq

\noindent where the first factor is the initial state distribution $\pi(\s_0)$, the second is the product of jump probabilities, the third integrates over waiting times, and the fourth constrains the total waiting time to be less than $t$ ($\Theta$ is the Heaviside step function).  In the limit $t \arr \infty$ we obtain the probability of the path $\varphi$ for \emph{any} duration,

\beq
\Pi_\infty[\varphi] = \pi(\s_0) \prod_{i=0}^{\ell-1} \me{\s_{i+1}}{\Q}{\s_i},
\label{eq:path_prob}
\eeq

\noindent which is just the product of jump probabilities.  In the time-dependent case, the Laplace transform of Eq.~\ref{eq:full_path_prob} results in a simpler expression due to deconvolution~\cite{Flomenbom2005}:

\beq
\tilde{\Pi}[\varphi, s] = \frac{\pi(\s_0)}{s} \prod_{i=0}^{\ell-1} \me{\s_{i+1}}{\Q}{\s_i} ~\tilde{\psi}_{\s_i}(s),
\label{eq:ltransform}
\eeq

\noindent where $\tilde{\psi}_{\s_i}(s)$ is the Laplace transform of $\psi_{\s_i}(t)$.  For Markov processes, Eq.~\ref{eq:ltransform} becomes~\cite{Sun2006}

\beq
\tilde{\Pi}[\varphi, s] = \frac{\pi(\s_0)}{s} \prod_{i=0}^{\ell-1} \frac{\me{\s_{i+1}}{\Q}{\s_i}}{1 + s w(\s)}.
\eeq

%%%%%%%%%%%%%%%%%%%%%%%%%%%%%%%%%%%%%%%%%%%%%%%%%%

\subsection{Path ensemble averages}

    The distribution $\Pi[\varphi, t]$ in principle contains all statistical information on a set of paths.  However, direct analysis of this distribution is typically prohibitive due to the high dimensionality of path space.  The simplest alternative entails taking averages of various path properties over this distribution.  Let $\Phi$ be an ensemble of paths that defines some dynamical process; for example, this may be all first-passage paths from a set of initial states $\mathcal{S}_i$ to a set of final states $\mathcal{S}_f$.  The partition function for this ensemble is
    
\beq
\Z_\Phi(t) = \sum_{\varphi \in \Phi} \Pi[\varphi, t],
\eeq

\noindent which represents the total probability of reaching $\mathcal{S}_f$ from $\mathcal{S}_i$ by time $t$.

    We define the following path functionals:

\begin{align}
\mathcal{L}[\varphi] & = \text{length of }\varphi & \mathcal{I}_\s[\varphi] & = \left\{ \begin{array}{l} 1\text{ if }\s \in \varphi \\ 0\text{ otherwise} \\ \end{array} \right. \nonumber \\
\mathcal{T}[\varphi] & = \sum_{i=0}^{\ell-1} w(\s_i) & \mathcal{T}_\s[\varphi] & = \sum_{i=0}^{\ell-1} \delta_{\s,\s_i} w(\s_i)
\end{align}

\noindent We can now express various path statistics as averages of these functionals over the ensemble, conditioned on completing the process by time $t$.  For example, the average time of paths is given by~\cite{Harland2007}

\beq
\bar{\tau}_\Phi(t) = \average{\mathcal{T}(t)}_\Phi = \frac{1}{\Z_\Phi(t)} \sum_{\varphi \in \Phi} \mathcal{T}[\varphi] \Pi[\varphi, t].
\eeq

\noindent The distribution of path lengths is given by

\beq
\rho_\Phi(\ell, t) = \frac{1}{\Z_\Phi(t)} \sum_{\varphi \in \Phi} \delta(\ell - \mathcal{L}[\varphi]) \Pi[\varphi, t],
\eeq

\noindent from which the average length $\bar{\ell}_\Phi(t) = \average{\mathcal{L}(t)}_\Phi$ and standard deviation of length $\ell_\Phi^\text{sd}(t)$ are readily obtained.

     Averages over state-dependent functionals can be used to characterize the spatial structure of paths.  For example, the fraction of time paths spend in a state $\s$ can be expressed as $\average{\mathcal{T}_\s(t)}_\Phi/\bar{\tau}_\Phi(t)$; this is a normalized distribution over states $\s$ and therefore can be considered as a density of states on paths in the ensemble $\Phi$.  In contrast, the probability that a path will hit a state $\s$ is given by $\average{\mathcal{I}_\s(t)}_\Phi$.  We will refer to this as the density of paths in the ensemble $\Phi$.  We can also express the two-point correlation function $\average{\mathcal{I}_{\s'}(t) \mathcal{I}_{\s}(t)}_\Phi$, which gives the probability of paths passing through both $\s'$ and $\s$. 

     In many cases we are interested in the time-independent versions of these quantities, i.e., we focus on statistical properties of paths taking any amount of time to finish.  These can be obtained as the $t \arr \infty$ limit of the above expressions, which amounts to replacing $\Pi[\varphi, t]$ (Eq.~\ref{eq:full_path_prob}) with $\Pi_\infty[\varphi]$ (Eq.~\ref{eq:path_prob}).  We will denote these time-independent properties by simply omitting the time dependence, e.g., $\lim_{t\arr\infty} \bar{\tau}_\Phi(t) = \bar{\tau}_\Phi$.

     Our formalism also allows for development of path thermodynamics.  The entropy of the path ensemble is given by

\beq
\begin{split}
S_\Phi(t) = & - \frac{1}{\Z_\Phi(t)} \sum_{\varphi \in \Phi} \Pi[\varphi, t] \log\left( \frac{\Pi[\varphi, t]}{\Z_\Phi(t)} \right) \\
= & - \average{\log \Pi(t)}_\Phi + \log \Z_\Phi(t).
\end{split}
\eeq

\noindent Indeed, if we consider the path Hamiltonian

\beq
H[\varphi, t] = - \log(\Pi[\varphi, t]),
\eeq

\noindent we can express the path ensemble free energy as

\beq
F_\Phi(t) = \average{H(t)}_\Phi - S_\Phi(t) = - \log \Z_\Phi(t).
\eeq

\noindent The partition function $\Z_\Phi(t)$ monotonically increases with time.  Therefore the free energy $F_\Phi(t)$ monotonically decreases as $t \arr \infty$, corresponding to equilibration of the path ensemble.

     For recurrent processes (i.e., those in which the system will almost surely reach the final states eventually~\cite{Redner2001}), $\lim_{t\arr\infty} \Z_\Phi(t) = \Z_\Phi = 1$, and hence equilibrium free energy is zero.  In these cases, equilibrium path entropy is equal to the average energy. If the ensemble $\Phi$ consists of only a single path with nonzero probability, its entropy is $S_\Phi = 0$.  This situation may arise if a landscape is so constrained that only a single viable pathway exists between the initial and final states.  Conversely, consider a purely random walk on a homogeneous network with $\gamma$ nearest neighbors per node.  The jump probability between any pair of neighboring nodes is thus $\bar{q} = 1/\gamma$, so any path $\varphi$ has probability $\Pi_\infty[\varphi] = \bar{q}^{\mathcal{L}[\varphi]}$, and the entropy of the ensemble is given by

\beq
S_\Phi = - \average{\log \Pi_\infty}_\Phi = - \bar{\ell}_\Phi \log \bar{q}.
\label{sphi:estimate}
\eeq

\noindent Note that path entropy and path energy scale with the average path length, which defines a notion of extensivity in the path ensemble.  This is sensible if we think of a path as a gas of particles, where each jump in the path corresponds to a particle.  The path ensemble, which includes paths of many lengths, therefore is equivalent to the grand canonical ensemble of the gas.  In the case of the gas, extensive quantities like entropy and energy scale with the number of particles, and hence these quantities here scale with the path length. In the case of evolutionary dynamics on fitness landscapes, $\bar{\ell}_\Phi \sim k^L$ and $\bar{q} \sim (L(k-1))^{-1}$ (where $L$ is the sequence length and $k$ is the alphabet size), yielding % MM: where does \bar{\ell}_\Phi come from??

\beq
S_\Phi \sim k^L \log L(k-1).
\eeq

%%%%%%%%%%%%%%%%%%%%%%%%%%%%%%%%%%%%%%%%%%%%%%%%%%

\subsection{Numerical algorithm}

     The factorized form of the path probability distribution functional (Eqs.~\ref{eq:full_path_prob}, \ref{eq:path_prob}) permits efficient calculation of path ensemble averages via a recursive algorithm.  Here for simplicity we consider the time-independent case.  Let $\ket{\pi}$ be the vector of initial state probabilities and $\ket{\s}$ be the vector with $1$ at position $\s$ and $0$ otherwise.  For each step $\ell$ and intermediate state $\s$, we can recursively calculate $P_\ell(\s) = \me{\s}{\Q^\ell}{\pi}$ and $T_\ell(\s)$, the total probability and average time of all paths that end at $\s$ in $\ell$ steps:
     
\begin{eqnarray}
%\begin{split}
P_{\ell}(\s') &=& \sum_{\text{nn } \s \text{ of } \s'} \me{\s'}{\Q}{\s} P_{\ell - 1} (\s), \\
T_{\ell}(\s') &=& \sum_{\text{nn } \s \text{ of } \s'} \me{\s'}{\Q}{\s} \left[ T_{\ell - 1} (\s) + w(\s) P_{\ell - 1} (\s) \right], \nonumber
%\end{split}
\end{eqnarray}

\noindent where $P_{0} (\s) = \pi (\s)$, $T_{0} (\s) = 0$, and the sums run over all nn $\s$ of $\s'$ ($\s' \in \mathcal{S}_f$ are treated as absorbing states).  This procedure generalizes the exact-enumeration algorithm of Ref.~\citep{Majid1984}.
Therefore $\Z_\Phi = \sum_{\ell=1}^\infty \sum_{\s \in \mathcal{S}_f} P_\ell(\s)$ and

\beq
\rho_\Phi(\ell) = \frac{1}{\Z_\Phi} \sum_{\s \in \mathcal{S}_f} P_\ell(\s), \quad \bar{\tau}_\Phi = \frac{1}{\Z_\Phi} \sum_{\ell=1}^\infty \sum_{\s \in \mathcal{S}_f} T_\ell(\s).
\eeq

\noindent Other ensemble averages such as $S_\Phi$, $\average{\mathcal{I}_\s}_\Phi$, $\average{\mathcal{I}_\s \mathcal{I}_{\s'}}_\Phi$, and $\average{\mathcal{T}_\s}_\Phi$ can be calculated similarly. Furthermore, we can calculate mean path divergence that characterizes the spatial diversity of the paths in $\Phi$:

\beq
\mathcal{D}_\Phi = \sum_{\ell = 0}^\infty \sum_{\s,\s' \in \mathcal{S}} d(\s,\s') P_\ell(\s) P_\ell(\s'),
\label{eq:mpd}
\eeq

\noindent where $d(\s, \s')$ is a distance metric on $\mathcal{S}$. Our definition is distinct from that proposed in Refs.~\citep{Lobkovsky2011, Lobkovsky2012} (Eq.~\ref{eq:mpd_Koonin}) in that it dynamically calculates distances between points on paths as they propagate, rather than comparing the minimal distance between complete paths.  Thus for a path that revisits some states multiple times, the divergence with a path
that travels through the same set of states without revisiting any of them will be zero according to Eq.~\ref{eq:mpd_Koonin}, but nonzero with the definition in Eq.~\ref{eq:mpd}.

     Our algorithm allows for very general definitions of the path ensemble $\Phi$ without having to explicitly enumerate all the paths. For instance, $\Phi$ can include paths that begin and end at arbitrary sets of states, or are not allowed to pass through arbitrary sets of intermediate states. Restriction to first-passage paths is also straightforward.  The time complexity of the algorithm is $\mathcal{O}(\gamma N\Lambda)$ ($\mathcal{O}(\gamma N^2\Lambda)$ for $\mathcal{D}_\Phi$), where $\gamma$ is the average number of nn defined above, $N$ is the number of states visited by paths in $\Phi$, and $\Lambda \sim \bar{\ell}_\Phi$ is the cutoff path length. For simple random walks, $\bar{\ell}_\Phi \sim N^{d_w/d_f}$ for $d_w > d_f$ and $\bar{\ell}_\Phi \sim N$ for $d_w \leq d_f$, where $d_w$ is the dimension of the walk and $d_f$ is the fractal dimension of the space~\cite{Bollt2005, Condamin2007}. Therefore, the algorithm scales as $\mathcal{O}(\gamma N^{1 + d_w/d_f})$ for $d_w > d_f$ and $\mathcal{O}(\gamma N^2)$ for $d_w \leq d_f$, automatically accounting for the sparseness of network connections.  This scaling compares favorably with standard linear algebra algorithms, which in general require $\mathcal{O}(N^3)$ operations~\cite{Press1992} to solve the backward equation~\cite{Metzner2009, Noe2009}.
 
\begin{figure}[t!]
\begin{center}
\begin{tabular}{ll}
\includegraphics[scale=0.35]{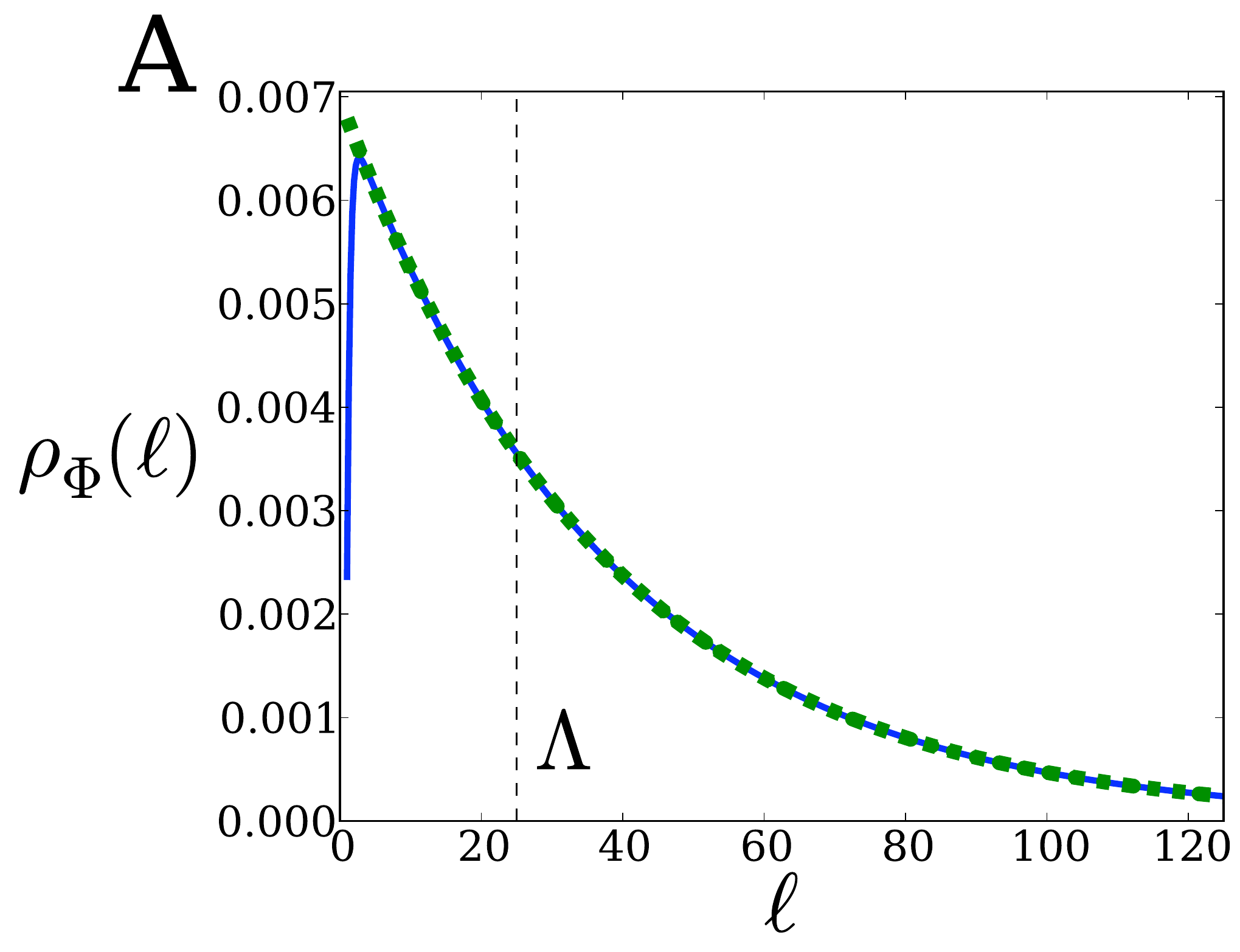} & \includegraphics[scale=0.35]{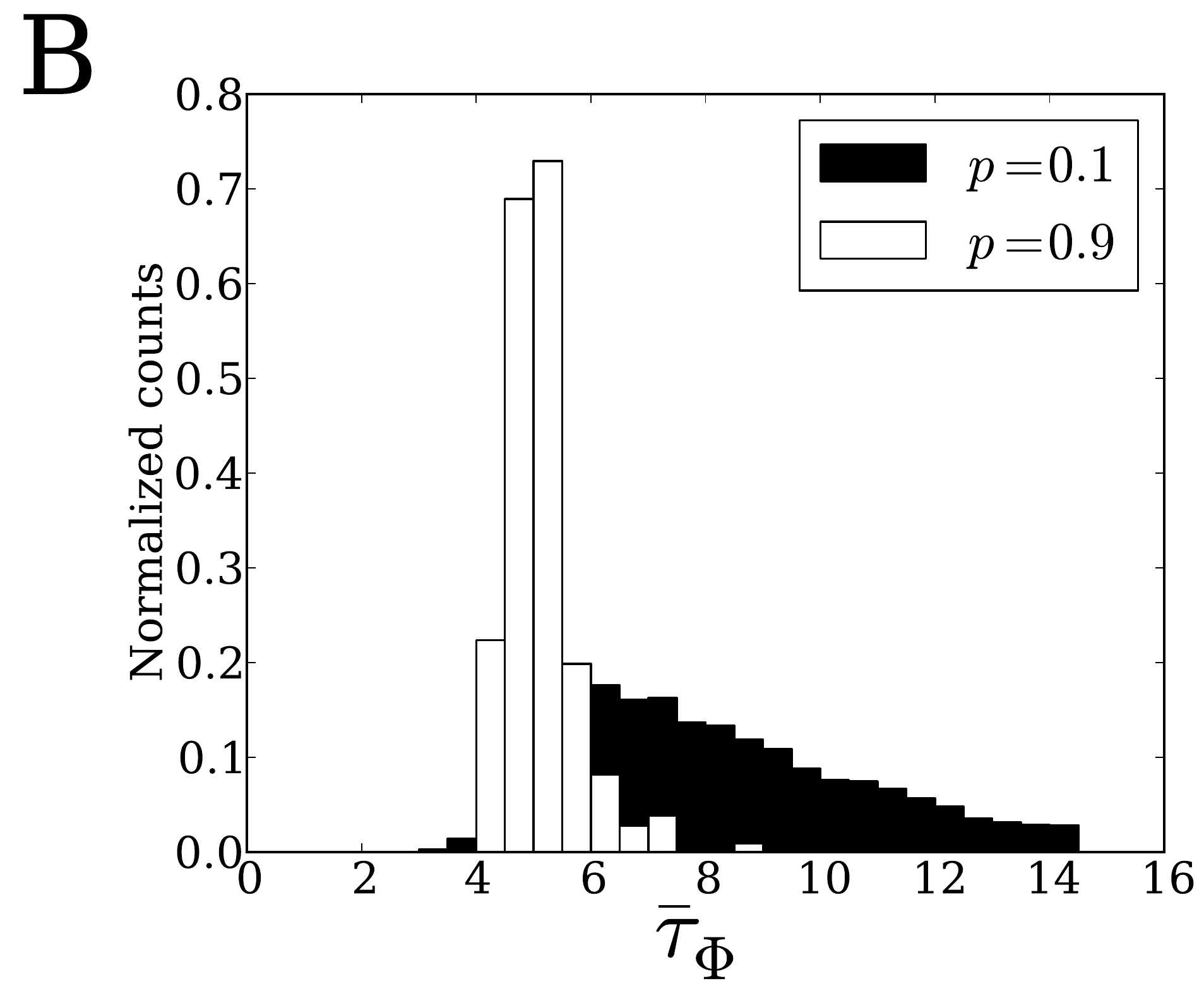} \\
\hspace{0.35cm} \includegraphics[scale=0.33]{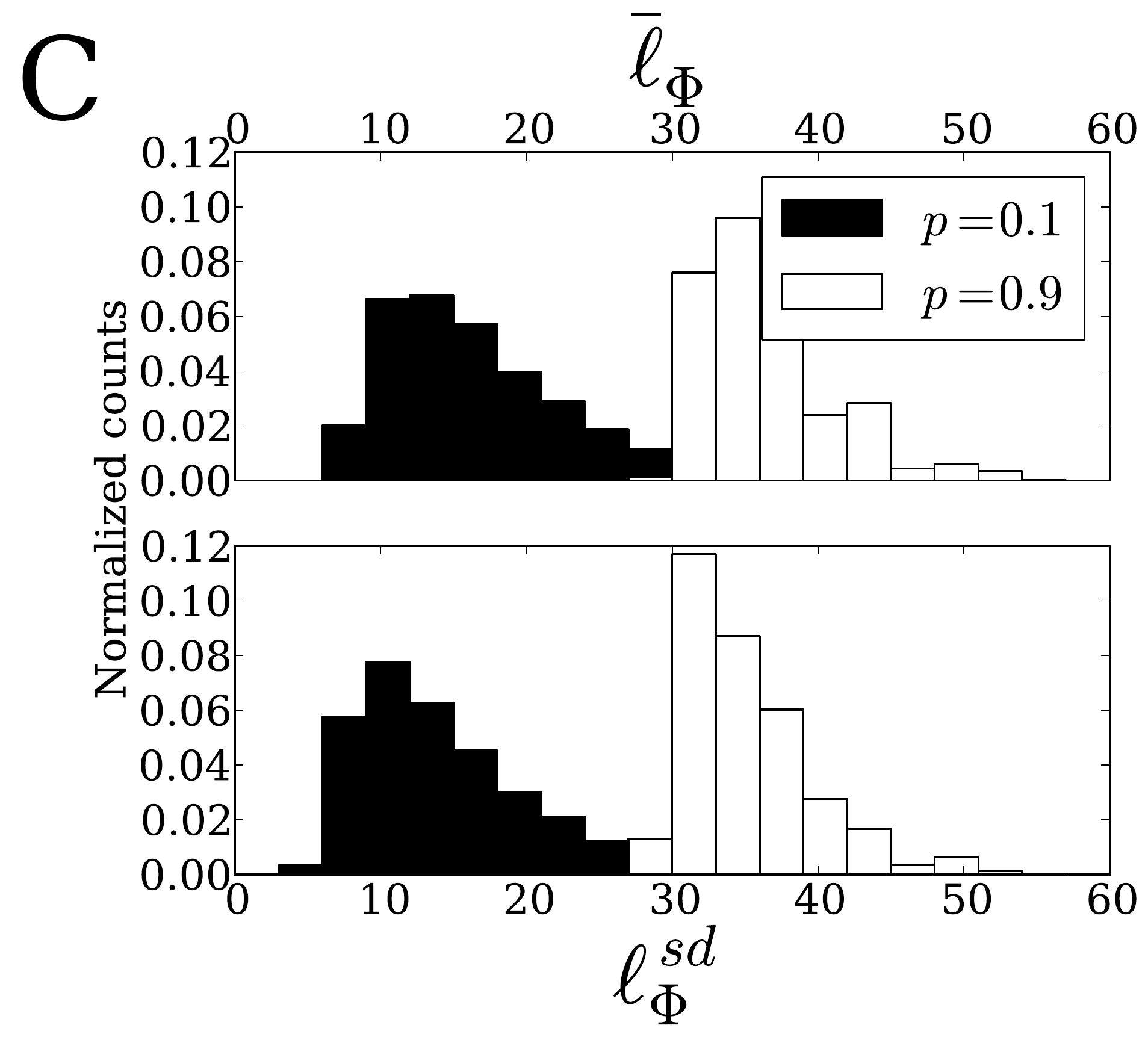} & \includegraphics[scale=0.35]{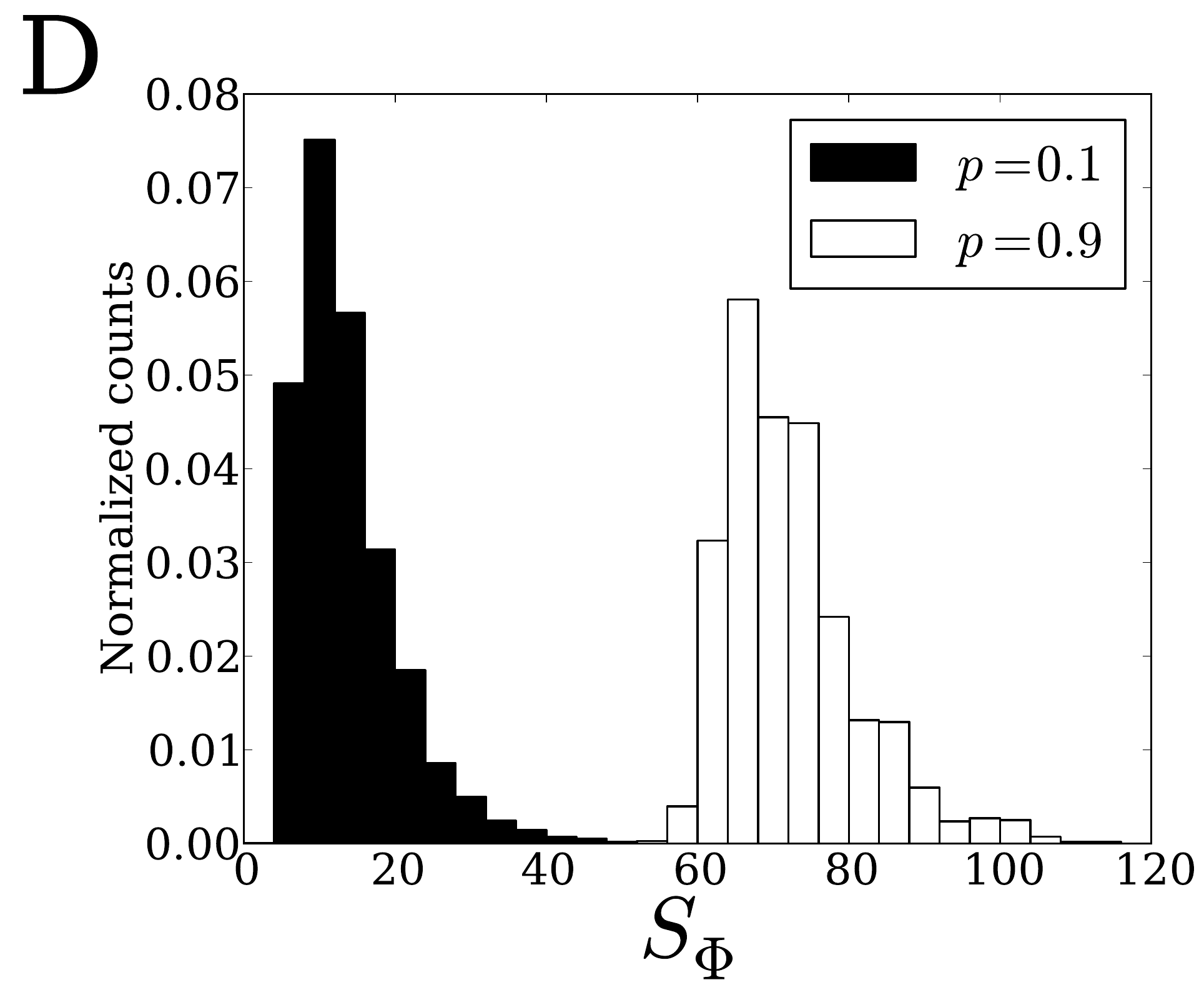} \\
\end{tabular}
\end{center}
\caption{Path ensemble statistics in a neutral network model. (A) The path length distribution $\rho_\Phi(\ell)$ (solid, blue) and exponential fit (dashed, green) in the interval $[\Lambda-5, \Lambda]$ for $\Lambda=25$ for a single realization of a neutral network with $p=0.9$. (B) Distribution of mean times of paths, (C) distribution of mean path lengths $\bar{\ell}_\Phi$ and standard deviations of path lengths $\ell_\Phi^\text{sd}$, and (D) distribution of path entropies $S_\Phi$ for $p=0.1$ and $p=0.9$.  All quantities in A-D are per site.  Histograms in B-D are generated from $10^4$ successful random realizations of the neutral network for each value of $p$; a realization is considered successful if both initial and final states are included in a single connected subnetwork.}
\label{fig:neutral_network}
\end{figure}

%%%%%%%%%%%%%%%%%%%%%%%%%%%%%%%%%%%%%%%%%%%%%%%%%%

\subsection{Evolution on a neutral network}

     As a simple application of our approach, we consider a population evolving on a neutral network~\cite{Franke2011}.  In the space of all sequences of length $L$ and with an alphabet of size $k$, we assign each sequence fitness 1 with probability $p$ or fitness 0 with probability $1-p$.  The subset of fit states connected to each other forms a neutral network; there can be several disconnected neutral subnetworks in each landscape realization.  We choose $L=8$ and a binary alphabet $\{\mathsf{A}, \mathsf{B}\}$ ($k=2$), which gives $2^8 = 256$ fit and unfit nodes in the network, and we consider the ensemble of first-passage paths from the sequence $\mathsf{AAAAAAAA}$ to the sequence $\mathsf{BBBBBBBB}$.  

     Figure~\ref{fig:neutral_network}A shows $\rho_\Phi(\ell)$ for a single realization of this model with $p=0.9$.  The exponential tail of $\rho_\Phi(\ell)$ is a universal feature of first-passage processes on finite spaces~\cite{Bollt2005}; other path statistics, such as the average time $\bar{\tau}_\Phi(\ell)$ of paths up to length $\ell$, also show asymptotic behavior that is exponential for long paths.  Indeed, we can use this feature to determine the cutoff path length $\Lambda$ for the algorithm: one need only consider paths with $\ell < \Lambda$ and infer the contributions of all longer paths from an exponential fit to the tail, which considerably improves the efficiency of the algorithm. This procedure takes advantage of the fact that information about longer paths is already contained in the structure of shorter paths; the maximum length $\Lambda$ of the shorter paths that must be calculated directly depends on the chemical distance between the initial and final states and the lengths over which the landscape is correlated.  This essentially implements a numerical renormalization scheme on the ensemble of paths~\cite{Sun2006}.
     
     In Fig.~\ref{fig:neutral_network}B,C,D we show distributions of the mean path time $\bar{\tau}_\Phi$, mean path length $\bar{\ell}_\Phi$, path length standard deviation $\ell_\Phi^\text{sd}$, and path entropy $S_\Phi$ for multiple realizations of the neutral network with high and low values of $p$.  We see that long paths are likely in these models; dozens of substitutions can occur at each site before the final state is reached.  The larger size of the neutral network for $p=0.9$ allows longer paths on average than for $p=0.1$.  However, the mean time of paths for the larger neutral network is usually smaller (Fig.~\ref{fig:neutral_network}B); this is because the increased connectivity of the network leads to shorter waiting times at individual nodes. Larger $p$ leads to substantially more diversity of paths and path lengths, as expected due to the increased size and connectivity of the network (Fig.~\ref{fig:neutral_network}C,D). Moreover, with $p=0.9$ the neutral network is nearly the size of the entire sequence space, and thus $\bar{q} \approx L^{-1}$ in Eq.~\ref{sphi:estimate}, leading to a factor of $\log L \approx 2.1$ difference between $\bar{\ell}_\Phi$ and $S_\Phi$ (Fig.~\ref{fig:neutral_network}C,D). Finally, note that the distributions of $\bar{\ell}_\Phi$ and $\ell_\Phi^\text{sd}$ in Fig.~\ref{fig:neutral_network}C are nearly the same, owing to the nearly exponential distribution of $\rho_\Phi(\ell)$ in this model (cf. Fig.~\ref{fig:neutral_network}A).
     
%     This statistical physics approach allows one to characterize the diversity of an ensemble of paths in several ways.  In particular, the entropy $S_\Phi$ characterize overall diversity of the path ensemble and the standard deviation $\ell^\text{sd}_\Phi$ characterizes the diversity in path lengths.  In many cases these different measures of path diversity will be correlated with each other.
     
%%%%%%%%%%%%%%%%%%%%%%%%%%%%%%%%%%%%%%%%%%%%%%%%%%%%%%%%%%%%%%%%%%%%%%%%%%%%%%%%%%%%%%%%%%%%%%%%%%%%

\section{Biophysics of protein evolution}

     We now consider a more realistic example of protein evolution on a fitness landscape which depends on protein folding stability and energetics of intermolecular interactions~\cite{Finkelstein2002}.  Many recent studies have focused on how proteins evolve under the constraint of maintaining thermodynamic stability of their folded state\cite{Bloom2005, DePristo2005, Bloom2006, Zeldovich2007, Bloom2007a, Bloom2007b, Bloom2007c, Bershtein2008}.  Suppose that an organism encodes a particular protein, and that the protein contributes fitness $1$ if it is folded and $f_0 < 1$ if it is unfolded.  Then the total fitness averaged over all proteins in an organism is given by 
     
\beq
\mathcal{F}(E_f) = \frac{1 + f_0 e^{\beta E_f}}{1 + e^{\beta E_f}},
\label{eq:fd_fitness}
\eeq

\noindent where $E_f$ is the free energy of folding (i.e., the free energy difference between folded and unfolded states of the protein) and $\beta = 1.7$ (kcal/mol)$^{-1}$ is inverse room temperature.  Equation~\ref{eq:fd_fitness} thus states that more reliably-folding proteins confer commensurate fitness advantages; often it is assumed that unfolded proteins are completely nonfunctional, and so $f_0 = 0$.  Some studies simplify this further by assuming that the folding energy $E_f$ need only be below a particular threshold $E_f^\text{thr}$; below that threshold all proteins are adequately stable and equivalent in fitness.  Mathematically,

\beq
\mathcal{F}(E_f) = \Theta(E_f^\text{thr} - E_f),
\eeq

\noindent where $\Theta$ is the Heaviside step function.  This model is equivalent to the zero-temperature limit of Eq.~\ref{eq:fd_fitness}.  Similar approaches based on protein-DNA interaction energies have also been used to study evolution of gene regulation~\cite{Sengupta2002, Gerland2002, Berg2003, Berg2004, Lassig2007, Mustonen2008}.

     Here we consider a model that includes both protein stability and function,~\cite{Manhart2013} which we take to be the ability to bind a target such as an enzymatic substrate or another protein.  Let $E_b$ be the free energy of binding relative to the chemical potential of the target molecule, so that the probability of binding is $1/(1 + e^{\beta E_b})$. We assume that the protein contributes fitness $1$ if it both folds and binds, and $f_0 < 1$ otherwise~\cite{Mayer2007}. Then fitness averaged over all proteins in an organism is given by
      
\beq
\mathcal{F}(E_f, E_b) = \frac{1 + f_0(e^{\beta E_f} + e^{\beta E_b} + e^{\beta(E_f + E_b)})}{(1 + e^{\beta E_f})(1 + e^{\beta E_b})}.
\label{eq:fitness}
\eeq

     The folding and binding energies depend on the amino acid sequence $\s$.  Many proteins have only a small number of residues at the binding interface that contribute the majority of the binding affinity; these residues are known as the ``hotspot''~\cite{Clackson1995}.  We assume that there are $L$ such residues and that they make additive contributions to the total binding and folding free energies~\cite{Serrano1993}:
     
\beq
E_f(\s) = E_f^0 + \sum_{\mu = 1}^L \epsilon_f(\mu, \s^\mu), \quad E_b(\s) = E_b^0 + \sum_{\mu = 1}^L \epsilon_b(\mu, \s^\mu),
\eeq

\noindent where $E_f^0, E_b^0$ are overall offsets and $\epsilon_{f,b}(\mu, \s^\mu)$ is the energy contribution of amino acid $\s^\mu$ at position $\mu$.  The offset $E_f^0$ is a fixed contribution to the folding energy from the other residues in the protein, which we assume to be perfectly adapted.  We sample $\epsilon_f$'s from a Gaussian with mean 1.25 kcal/mol and standard deviation 1.6 kcal/mol, consistent with computational studies showing the mutational effects on stability are universally distributed~\cite{Tokuriki2007}.  Since binding hotspot residues are typically defined as those having a minimum penalty of 1--3 kcal/mol for mutations away from the wild-type amino acid~\cite{Bogan1998}, we set $\epsilon_b(\mu, \s^\mu_\text{bb}) = 0$ for all $\mu = 1, \ldots, L$ ($\s_\text{bb}$ is the best-binding sequence: $E_b (\s_\text{bb}) = E_b^0$), and we sample the other $\epsilon_b$'s from an exponential distribution defined in the range of $(1,\infty)$ kcal/mol, with mean 2 kcal/mol. The parameters of the exponential distribution are consistent with alanine-scanning experiments which probe energetics of amino acids at the binding interface~\cite{Thorn2001}.
By construction, $E_b^0$ is the binding free energy of the best-binding sequence, measured relative to the chemical potential of the target molecule.
Here we consider $L=5$ hotspot residues and a reduced alphabet of $k=8$ amino acids grouped by physico-chemical properties, resulting in $8^5 = 32768$ possible sequences.  Different choices of these parameters can be considered, but they appear to have little effect on the overall qualitative features of the model.

     %A choice of $E_f^0$ and $E_b^0$ and sets of $\epsilon_f$ and $\epsilon_b$, drawn from the aforementioned distributions, define a fitness function of sequence through Eq. \ref{eq:fitness}.

\begin{figure}[t!]
\begin{center}
\includegraphics[scale=0.35]{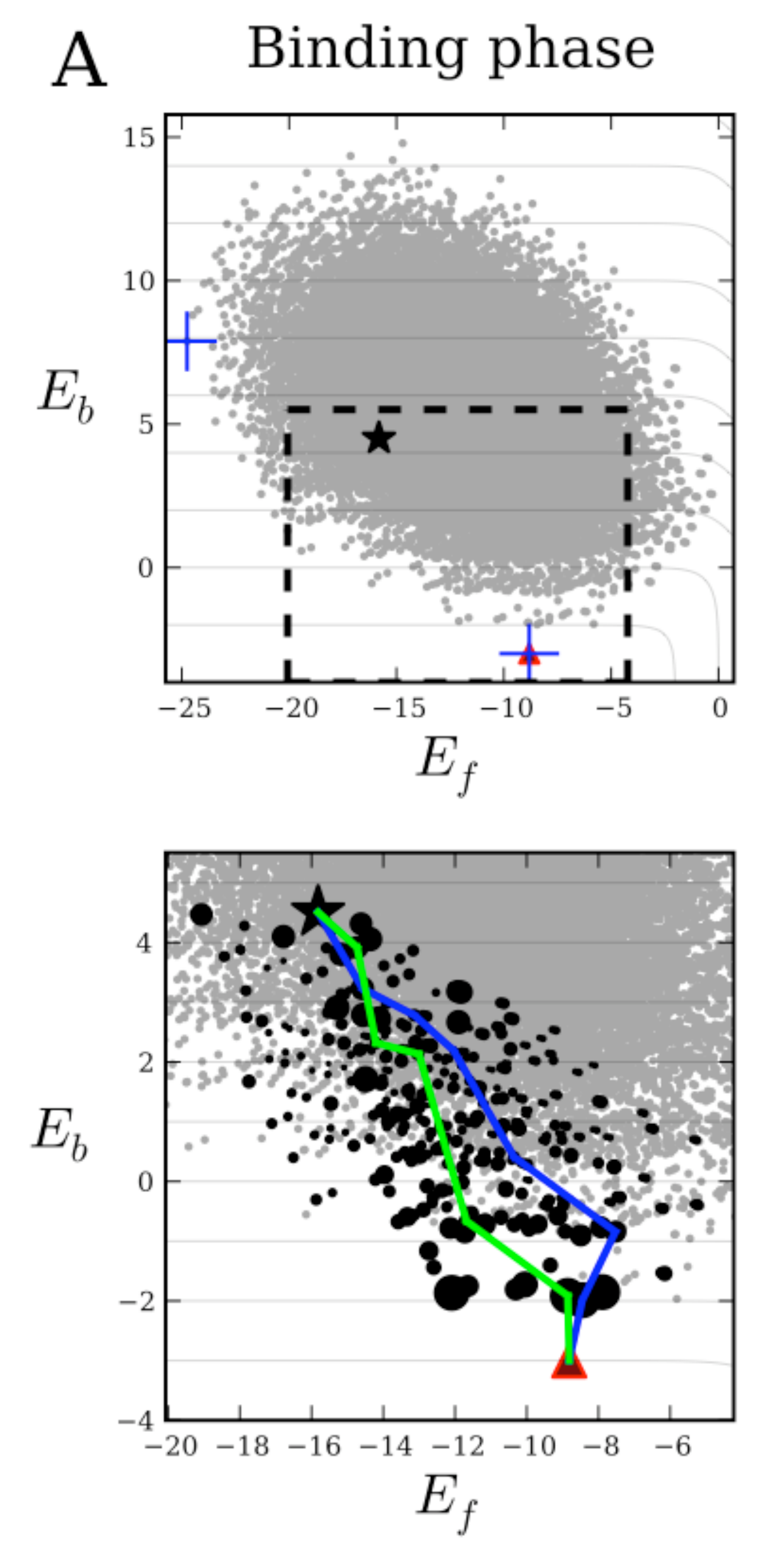}
\includegraphics[scale=0.35]{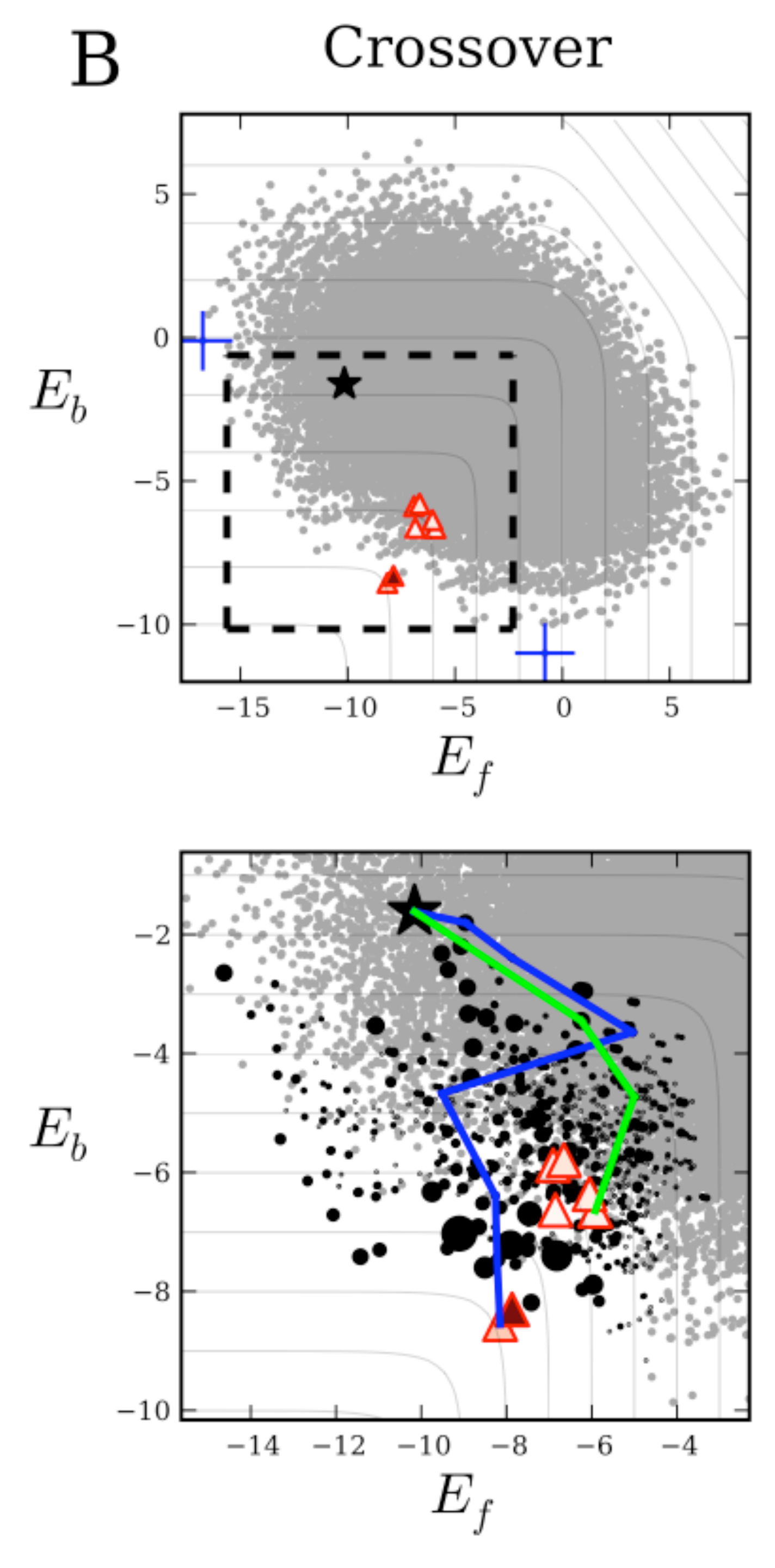}
\includegraphics[scale=0.35]{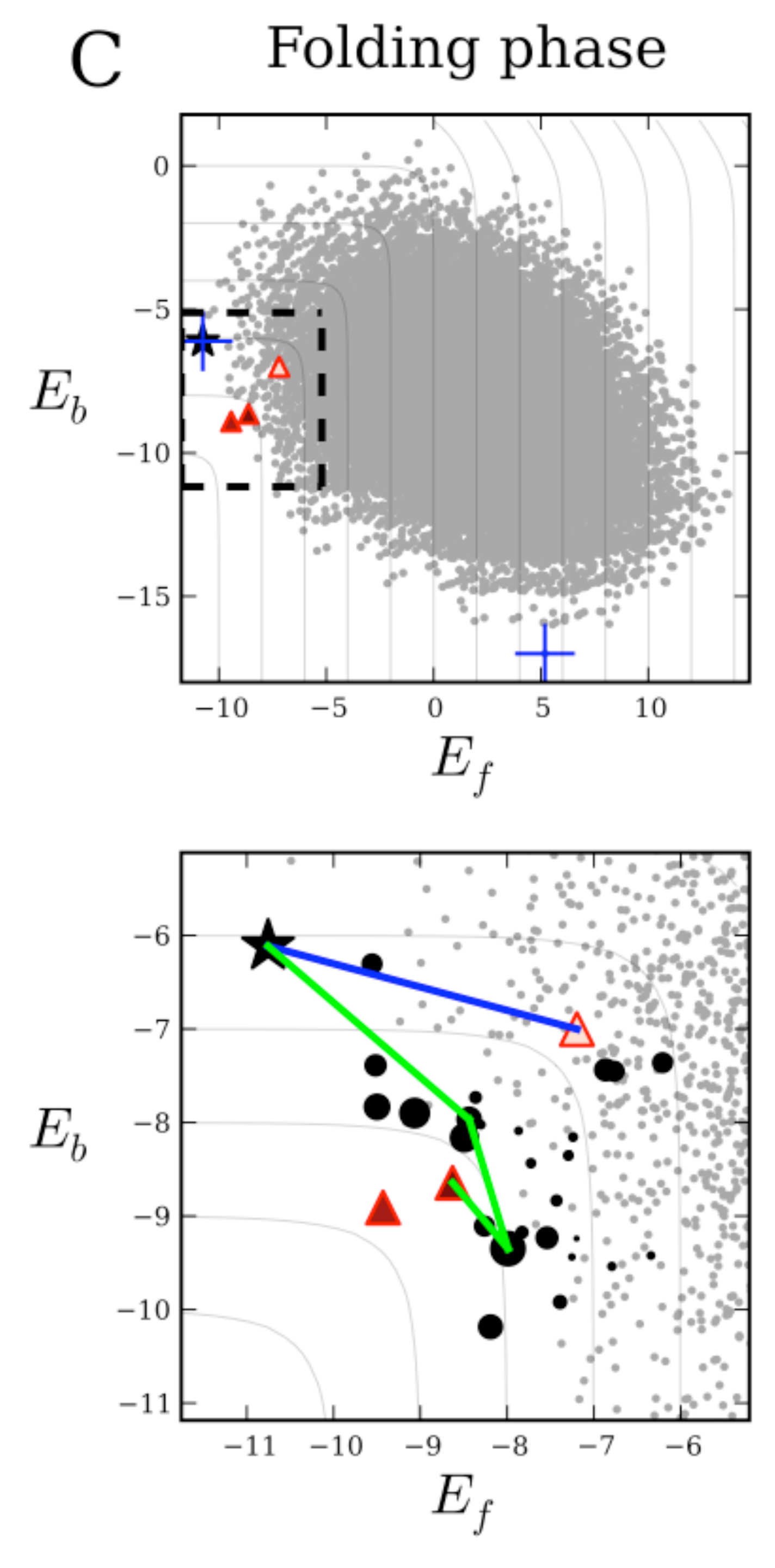}
\end{center}
\caption{Three realizations of the fitness landscape. The offsets $E_f^0$ and $E_b^0$ are different for each realization, but $\epsilon_f$'s and the two sets of $\epsilon_{b}$'s (one for $\mathcal{F}_1$ and another for $\mathcal{F}_2$) are the same.
(A)~Binding phase, with $E_f^0 = -17$~kcal/mol and $E_b^0 = -3$~kcal/mol.
(B)~Crossover regime, with $E_f^0 = -9$~kcal/mol and $E_b^0 = -11$~kcal/mol.
(C)~Folding phase, with $E_f^0 = -3$~kcal/mol and $E_b^0 = -17$~kcal/mol.
%($E_f (\s^\text{bf}) = -9.4$~kcal/mol).
Top panels of A-C show the global distribution of all $8^5 = 32768$ sequences in energy space according to $\mathcal{F}_2$, where the blue crosses indicate the best-folding ($\s_\text{bf}$) and best-binding ($\s_\text{bb}$) sequences, red triangles indicate local fitness maxima on $\mathcal{F}_2$ (shaded according to their commitment probabilities), and black stars indicate the initial state for adaptation (sequence with global maximum on $\mathcal{F}_1$).  Black lines are contours of constant fitness $\mathcal{F}_2$.  In the bottom panels of A-C only the region of energy space accessible to APs is shown; this region is outlined by dashed lines in the top panels. Example APs are shown in blue and green, and black circles indicate intermediate states along APs, sized proportional to the AP density $\average{\mathcal{I}_\s}_\AP$.}
\label{fig:landscapes}
\end{figure}
 
     We consider a population of individuals whose genomes encode a protein of interest. In each individual, the sequence of the protein begins as perfectly adapted to binding a certain target molecule.  Then the population is subjected to a selection pressure that favors binding to a new target. This situation is common in directed evolution experiments which attempt to evolve new protein functions in laboratory settings~\cite{Bloom2009}. To model such experiments, we sample one set of $\epsilon_f$'s and two sets of $\epsilon_{b}$'s (one for each target), while $E_f^0$ and $E_b^0$ are assumed to be fixed.  This procedure defines two fitness landscapes, $\mathcal{F}_1$ and $\mathcal{F}_2$, through Eq.~\ref{eq:fitness}; the entire population begins at the global maximum on $\mathcal{F}_1$ and proceeds to adapt to a new global or local maximum on $\mathcal{F}_2$.  We assume the SSWM limit as described in the Introduction; for Markovian waiting times, the jump probabilities are thus $\me{\s'}{\Q}{\s}=1/b(\s)$ if $\mathcal{F}(\s') > \mathcal{F}(\s)$ and zero otherwise, where $b(\s)$ is the number of beneficial substitutions possible from $\s$.  Note that in this limit our results are independent of $f_0$ and the mutation rate and effective population size only affect the overall time scale. The path ensemble consists of all adaptive paths (APs); fitness monotonically increases along each path. In Fig.~\ref{fig:landscapes} we show three realizations of $\mathcal{F}_2$ with examples of APs.

\begin{figure}[t!]
\begin{center}
\includegraphics[scale=0.7]{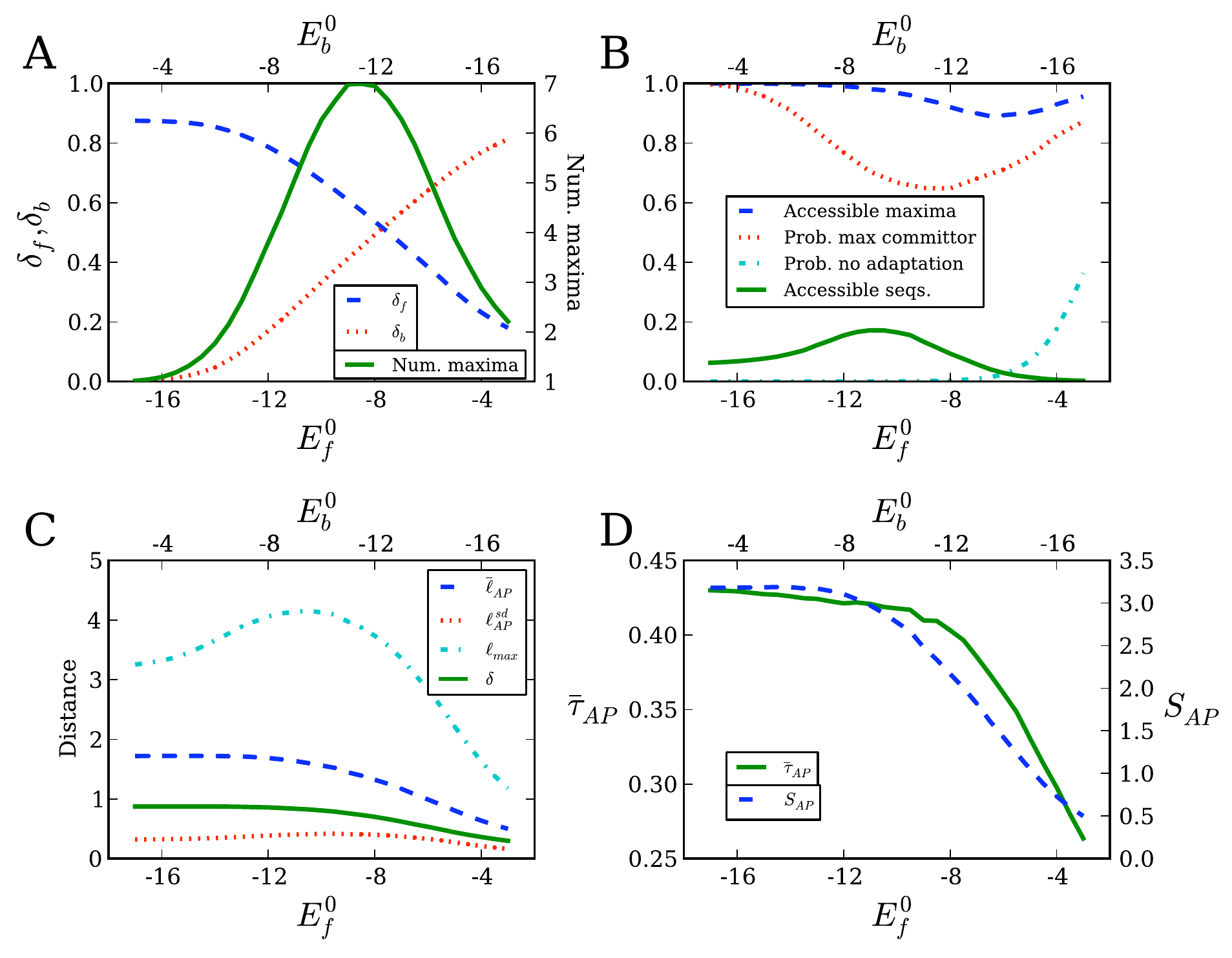}
\end{center}
\caption{
Statistics of fitness landscapes and dynamics averaged over multiple landscape realizations.
All quantities in A-D are per-residue (except number of local maxima).
(A)~Average number of local fitness maxima (solid, green), the average Hamming distance $\delta_f$ (number of mutations) between the maxima and the best-folding sequence $\s_\text{bf}$ (dashed, blue), and the average distance $\delta_b$ between the maxima and the best-binding sequence $\s_\text{bb}$ (dotted, red) for the parameter subspace $E_f^0 + E_b^0 = -20$~kcal/mol. Note that the average distance between two random sequences is $1 - 1/k = 0.875$, where $k=8$ is the size of the alphabet.
(B)~Fraction of local fitness maxima accessible from the initial state (dashed, blue), fraction of all landscape realizations in which the global maximum has the largest commitment probability among all local maxima (dotted, red), probability that the initial sequence starts at a local maximum resulting in no adaptation (dashed and dotted, cyan), and fraction of all sequences accessible to APs (solid, green).
(C)~Mean AP length $\bar{\ell}_\AP$, standard deviation $\ell_\AP^\text{sd}$ of APs,
average distance $\delta$ between the initial state and the final states, and average length $\ell_\text{max}$ of the longest APs connecting the initial state with the final states.
(D)~Path ensemble entropy $S_\AP$ (dashed, blue) and the mean duration of paths $\bar{\tau}_\AP$ (solid, green), in units of inverse population mutation rate $(Nu)^{-1}$.
The probability of no adaptation in (B) is an average over $2 \times 10^4$ landscape realizations; all other data points are averages over $5 \times 10^3$ realizations, and realizations with no adaptation are excluded.}
\label{fig:phases}
\end{figure}

     We focus on the generic properties of these fitness landscapes, averaged over many realizations of $\epsilon_f$ and $\epsilon_b$ (Fig.~\ref{fig:phases}). Varying $E_f^0$ and $E_b^0$ reveals two qualitatively different phases of adaptation.  When $E_f^0$ is low and $E_b^0$ is high (see Fig.~\ref{fig:landscapes}A for an example), adaptation is in the \emph{binding phase}, i.e., the need to bind the new target molecule dominates evolutionary dynamics. In this phase, there is typically a single local fitness maximum which coincides with the best-binding sequence $\s_\text{bb}$ (Fig.~\ref{fig:phases}A). In contrast, when $E_f^0$ is high and $E_b^0$ is low (see Fig.~\ref{fig:landscapes}C for an example), adaptation is in the \emph{folding phase}, where evolution is mostly constrained by the need to maintain and increase protein stability.  In this case there are also few local maxima and they tend to be close in sequence space to the best-folding sequence $\s_\text{bf}$ (Fig.~\ref{fig:phases}A).  Between these phases there is a crossover regime, where folding and binding compete more equally in shaping the landscape and adaptation (see Fig.~\ref{fig:landscapes}B for an example).  The crossover regime has the most epistasis, as indicated by the number of local maxima, the accessibility of those maxima, and the fraction of fitness landscape realizations in which the global maximum has the largest commitment probability (Fig.~\ref{fig:phases}A,B). The differences in the landscape structure in the binding and folding phases lead to substantial differences in adaptive dynamics. In particular, APs are longer and take more time in the binding phase compared to the folding phase; they are also more diverse (Fig.~\ref{fig:phases}C,D).
Initial and final states in the binding phase are separated by longer Hamming distances (Fig.~\ref{fig:phases}C).
In the folding phase, there is an appreciable probability that no adaptation occurs, since the initial state may coincide with one of the local maxima
(Fig.~\ref{fig:phases}B).

     Our model reproduces several important features of molecular evolution observed in experimental studies.  First of all, adaptive dynamics may involve tradeoffs between folding and binding frequently observed in directed evolution experiments~\cite{Tokuriki2009, Bloom2009, Wang2002}, even though mutational effects on folding and binding energies are uncorrelated~\cite{Tokuriki2008}.  This coupling between folding and binding is introduced through nonlinearities in the fitness function $\mathcal{F}(E_f, E_b)$ (Eq.~\ref{eq:fitness}), which contains both magnitude and sign epistasis. We note that although our landscapes are generated from randomly-drawn parameters $\epsilon_{f}$ and $\epsilon_{b}$, similar to many classical model landscapes, these these protein landscapes are highly correlated: fitness values of $k^L$ sequences are determined by $2Lk$ parameters.  Indeed, the average number of local maxima on a House of Cards landscape with $L=5$ and $k=8$ is $\approx 910$, while the protein landscape generally has no more than 7 (Fig.~\ref{fig:phases}A).  Thus, our models are less epistatic than completely uncorrelated landscapes~\cite{Kauffman1993}. These more moderate levels of epistasis are consistent with previous analyses of empirical fitness landscapes~\cite{Carneiro2010, Lobkovsky2011, Szendro2013}.  In the folding phase, APs tend to be short and no adaptation may occur if the old global maximum coincides with the new local maximum (Fig.~\ref{fig:phases}B). This lack of adaptation is sometimes observed in experiments in which a protein already exhibiting some affinity for the new ligand cannot readily increase it any further~\cite{Bloom2009}.  In the crossover regime, the tradeoff between binding and folding alone can result in proteins with marginal folding stability, in contrast with previous hypotheses that explain marginal stability with mutational entropy~\cite{Zeldovich2007} or a fitness function that disfavors hyperstable proteins~\cite{DePristo2005}.

\begin{figure}[t!]
\begin{center}
\includegraphics[scale=0.4]{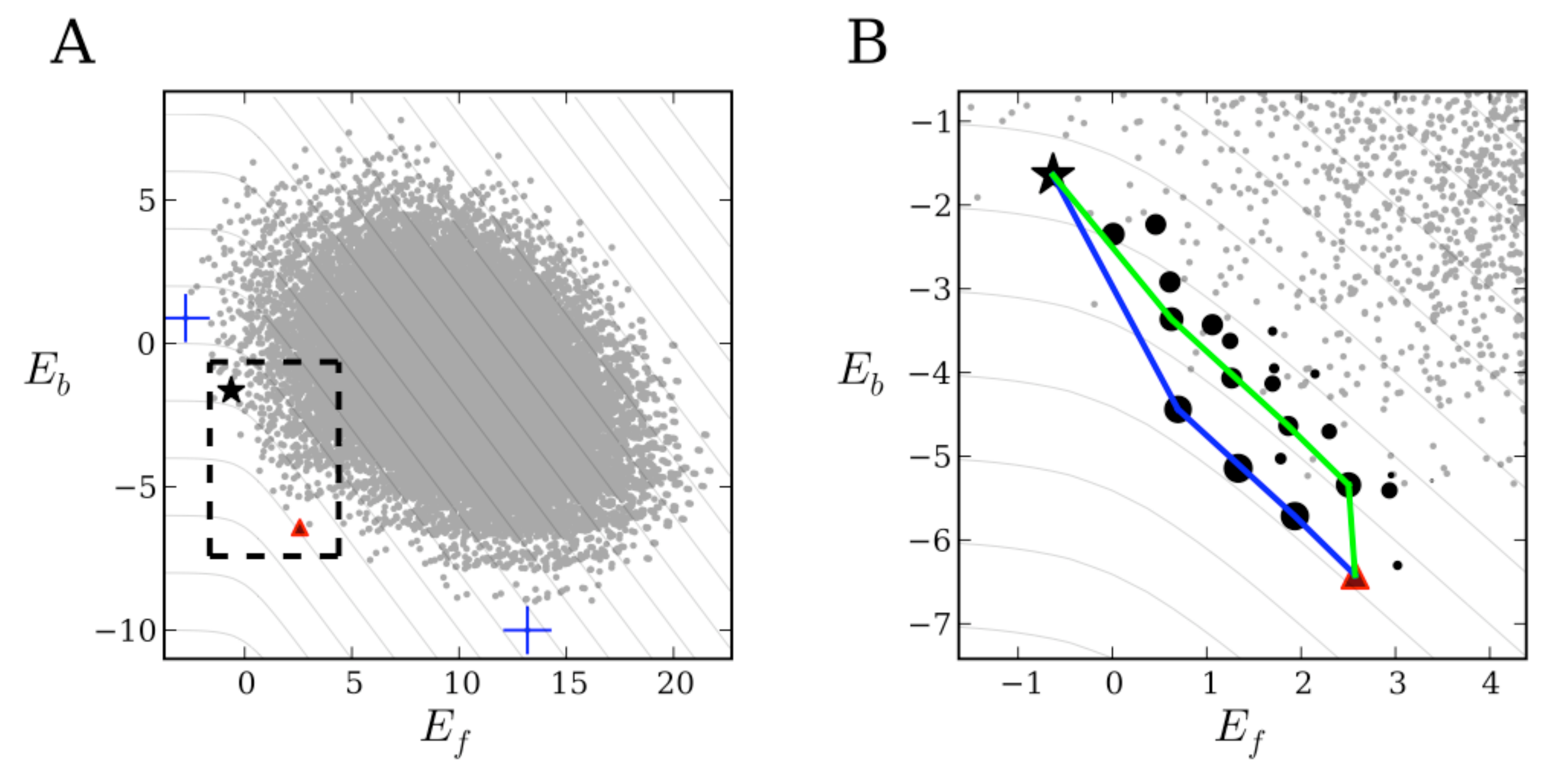}
\end{center}
\caption{A realization of the fitness landscape in Eq.~\ref{eq:3state_fitness}, with $E_f^0 = 5$ kcal/mol and $E_b^0 = -10$ kcal/mol.  Notation is the same as in Fig.~\ref{fig:landscapes}.}
\label{fig:3state}
\end{figure} 

     Our model can be extended to account for binding-mediated stability, in which binding stabilizes an otherwise disordered protein~\cite{Brown2011}.  In this case, instead of considering folding and binding to be independent as in Eq.~\ref{eq:fitness}, we assume that the protein can only bind when it is folded. The unbound protein may be biased toward the unfolded state, which effectively creates a free energy barrier between the ``unfolded
and unbound'' state and the globally most favorable ``folded and bound'' state.  The fitness function in this model is given by
     
\beq
\mathcal{F}(E_f, E_b) = \frac{1 + f_0(e^{\beta E_b} + e^{\beta(E_f + E_b)})}{1 + e^{\beta E_b} + e^{\beta(E_f + E_b)}}.
\label{eq:3state_fitness}
\eeq

\noindent Here, proteins may still have high fitness even if $E_f > 0$ as long as $E_f + E_b < 0$, as expected for a protein stabilized by binding.  Our methodology can be straightforwardly applied to this case as well, again revealing the existence of binding- and folding-dominated phases.  See Fig.~\ref{fig:3state} for an example landscape.
     
     We can also incorporate chaperone-assisted folding \cite{Rutherford2003} by modifying $E_f^0$ or the distribution of $\epsilon_f$'s, and include ``folding hotspots'' away from the binding interface, which may acquire stabilizing mutations as a buffer against destabilizing but function-improving mutations at the interface \cite{Tokuriki2009, Bloom2009}. Neutral and slightly deleterious mutations can be incorporated as well by using substitution rates from more complex population genetics models~\cite{Crow1970, Manhart2012}, although we expect non-adaptive substitutions to play little role on short time scales.
     
	In summary, we have described a general methodology for studying stochastic paths on arbitrary landscapes and networks.
%based on recursive updates of key statistics of the complete path ensemble.
This approach is general and can be applied to numerous dynamical problems in physics, chemistry, biology, and engineering, including protein folding, transport and search in complex media, stochastic phenotypes, and cell-type differentiation.  In this review we have emphasized its utility in exploring evolutionary problems, which can often be visualized as random walks on fitness landscapes~\cite{Wright1932} and where the diversity and reproducibility of evolutionary paths is a central issue.  We believe that the path-based methodology is well-suited for providing intuitive path statistics in problems whose complexity and high dimensionality make direct visualisations impossible.

%%%%%%%%%%%%%%%%%%%%%%%%%%%%%%%%%%%%%%%%%%%%%%%%%%%%%%%%%%%%%%%%%%%%%%%%%%%%%%%%%%%%%%%%%%%%%%%%%%%%

\bibliographystyle{unsrtnat}
\bibliography{Chapter}

\end{document}